# Broadband Dielectric and THz Spectroscopy on Bio-Related Matter: Water, Amino Acids, Proteins, and Blood

Peter Lunkenheimer, Sebastian Emmert, Martin Wolf, and Alois Loidl

**Abstract** In the present work, we examine the relevance and proper interpretation of broadband-dielectric and THz-spectroscopy data for the investigation of various types of biological matter. We provide an overview of the rich variety of different dynamic processes that can be detected by these experimental methods. Several experimental examples are discussed in detail, helping to understand the information that can be drawn from such studies. This includes dielectric spectra, extending well into the GHz region, for pure water, which can be considered as a simple but highly important biological molecule. We also discuss results for a prototypical aqueous solution of a protein, belonging to one of the most important classes of biological macromolecules. Moreover, we examine broadband dielectric spectra on blood as an example of functional biological matter in organisms. To demonstrate the relevance of THz spectroscopy for the investigation of biological molecules, we finally treat such experiments applied to different amino acids.

———————————

P. Lunkenheimer (✉), S. Emmert, M. Wolf, A. Loidl

Experimental Physics V, Center for Electronic Correlations and Magnetism, University of Augsburg, 86159 Augsburg, Germany

e-mail: peter.lunkenheimer@physik.uni-augsburg.de

# 1 Introduction

Most biological systems reveal a rich variety of dynamic processes of their microscopic constituents, which often impact their vital functions in organisms, are relevant for electromagnetic dosimetry, and can be employed for diagnostic and therapeutic applications [1,2,3,4,5,6,7]. This includes the relatively slow inter- and intramolecular motions of large biomolecules like proteins, the translational hopping of dissolved ions, interfacial effects, e.g., due to the presence of cell membranes, the motions of bound and free water molecules, phonon-like excitations, the stretching and bending of intramolecular bonds, any many more. These processes occur on partly very different time scales, extending from seconds or slower, down to picoseconds. When investigating them by spectroscopic methods, each process ideally should lead to a separate spectral feature in the detected susceptibilities, which can show up at frequencies ranging from sub-Hz to THz, depending on the particular type of motion. Thus, performing spectroscopy covering a broad frequency range is crucial for the investigation of the dynamics in biological systems. This requirement is most easily achieved by broadband dielectric measurements, complemented by THz experiments. Fortunately, most of the dynamic processes in biological systems involve dipolar fluctuations or charge transport, which is prerequisite for their investigation by these methods. Indeed, the application of dielectric and THz spectroscopy to biological matter is well establish since many decades [1,2,3,4,5,6,7,8]. The determination of the dielectric properties and the achievement of a better understanding of the dynamic processes in biological systems is highly relevant in many respects. Important examples are the fixing of limiting radiation values to control electromagnetic pollution, the development of new diagnostic and therapeutic techniques, and food monitoring [3,6].

In the present work, we first provide a brief overview of the most important dynamic processes revealed in the dielectric spectra of biological matter. We then discuss the dielectric response of pure water, which governs the ubiquitous $\gamma$ relaxation found in most biological materials, and treat the broadband dielectric spectra of aqueous protein solutions. We further consider results on blood as an example for the complex dielectric behavior of functional biological matter in organisms and, finally, discuss the vibrational modes in amino acids as detected by THz spectroscopy. Concerning experimental methods and dielectric quantities, we refer the reader to the numerous earlier publications treating these topics, e.g., Refs. [1,2,3,6,9,10,11].

# 2 Dielectric Properties of Biological Matter

## *2.1 Overview*

Fig. 1 shows a schematic dielectric-loss spectrum, $\varepsilon''(\nu)$, indicating several typical contributions arising in biological materials. Most of the latter include water with dissolved mobile ions, which can lead to considerable dc conductivity. As explained in section 2.2, this generates a $1/\nu$ divergence in $\varepsilon''(\nu)$ at low frequencies ($\nu$ denotes the frequency of the applied ac field). As indicated by the dashed line in Fig. 1, this contribution can superimpose considerable parts of the relaxational contributions to the loss spectra and often is subtracted to more clearly reveal them (solid line). Ionic conductivity usually causes electrode polarization (EP), leading to huge low-frequency contributions in the real part of the permittivity $\varepsilon'$, but it also affects $\varepsilon''$ (not shown in Fig. 1) as explained in section 2.3. The denotations $\beta$, $\gamma$, and $\delta$ for the relaxational processes, as included in Fig. 1, follow the nomenclature used in large parts of the biophysical literature treating dielectric properties [2,12,13] (the $\alpha$ relaxation detected in some materials at very low frequencies is not show as it usually is observed in $\varepsilon'$ only; see section 2.4). The underlying microscopic processes of these relaxations are discussed in sections 2.5 – 2.7. The abbreviation "res." in Fig. 1 stands for resonances due to vibrational excitations (section 2.8) and "BP" denotes the boson peak (section 2.9), which should be present in all disordered biological systems.

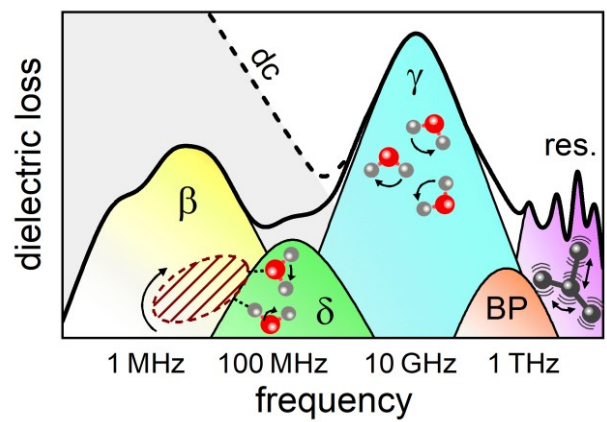

**Fig. 1** Schematic representation of some typical contributions to dielectric-loss spectra of biological matter (double-logarithmic representation). The dashed line shows the spectrum including the dc conductivity and the solid line indicates the behavior after its subtraction. The denotations $\beta$, $\gamma$, $\delta$ follow the biophysics nomenclature [2,12,13]. $\beta$ relaxations can be bimodal (visualized by the indicated double peak in the figure) and arise from reorientations of large dipolar entities (hatched ellipse in the figure) and/or Maxwell-Wagner relaxations due to internal heterogeneities (e.g., cell membranes). The $\delta$ relaxation also can be multimodal (not indicated) and mostly is ascribed to motions of bound water molecules and/or internal protein dynamics. The $\gamma$ relaxation reflects free water-molecule reorientations also occurring in bulk water. "BP" stands for the boson peak as suggested for unbound water molecules [48] and detected for other biological matter using scattering techniques [70]. Finally, "res." denotes high-frequency resonances, e.g., due to bond stretching and bending. The controversial $\alpha$ relaxation (section 2.4) and possible sub-$\beta$ processes due to conformational sampling [5] are not included in this figure.

To avoid confusion, it should be mentioned here that the naming of loss peaks by Greek letters is also common practice within the field of supercooled liquids and other glass-forming materials [14,15]. However, there their assignment is completely different and, for example, the biophysical $\beta$ relaxation arising from protein reorientations would be denoted as $\alpha$ relaxation.

## *2.2 Dc Conductivity*

Most dielectrically investigated biological matter such as tissues, cell suspensions (e.g., blood; see section 3.3), or solutions of biomolecules (e.g., proteins; see section 3.2) contain considerable amounts of water including dissolved ions. The high mobility of the latter in water [16] leads to considerable dc conductivity, $\sigma_{dc}$, i.e., biological samples usually are ionic conductors. There is a general relation between the dielectric loss $\varepsilon''$ and the real part of the conductivity $\sigma'$, namely: $\varepsilon'' = \sigma'/(2\pi \nu \varepsilon_0)$ (here, $\varepsilon_0$ is the permittivity of vacuum). Hence, in $\varepsilon''(\nu)$ the dc conductivity leads to a contribution $\varepsilon''_{dc} = \sigma_{dc}/(2\pi \nu \varepsilon_0)$, implying a $1/\nu$ divergence in the loss which obscures the detection of possible relaxation processes at low frequencies (cf. dotted line in Fig. 1). A subtraction of the dc contribution can uncover these relaxations as indicated by the solid line in Fig. 1 at low frequencies. However, at very low frequencies, the insufficient precision of the experimental data can lead to considerable scatter in the subtracted results. Thus, very slow dynamic processes (e.g., the $\alpha$ relaxation) are difficult to detect in $\varepsilon''(\nu)$ in this way. This problem does not apply to the real part of the permittivity, $\varepsilon'$, because the dc conductivity does not contribute to this quantity. However, in systems with high ionic conductivity, electrode polarization (section 2.3) usually dominates the $\varepsilon'$ spectra, also hampering the detection of intrinsic relaxation processes at low frequencies.

## *2.3 Electrode Polarization and Maxwell–Wagner Relaxations*

In typical dielectric experiments, the ionic charge transport present in most biological materials leads to the inevitable occurrence of so-called electrode polarization (EP) as observed in all ionic conductors [17,18]. It is caused by the accumulation of essentially immobile ions at the metallic sample electrodes, arising at low frequencies due to the simple fact that the ions cannot enter the metal. This leads to thin, poorly conducting layers close to the electrodes, an effect sometimes termed "blocking electrodes". They act as huge capacitors connected in series to the bulk sample. These effects lead to so-called Maxwell-Wagner relaxations causing a strong frequency dependence of the dielectric properties. Maxwell and Wagner [19,20] showed that in inhomogeneous materials, consisting of two or more regions with different dielectric properties, non-intrinsic dispersion effects in the permittivity and conductivity can arise, closely resembling the behavior of intrinsic dipolar relaxation processes [i.e., a steplike decrease in $\varepsilon'(\nu)$ and a peak in $\varepsilon''(\nu)$]. The special situation of thin, poorly conducting layers in series to the bulk sample can be most easily modelled by an equivalent circuit consisting of two parallel RC circuits (one for the blocking-electrode layers and one for the bulk) that are connected in series to each other [18,21,22]) This leads to the signature of a relaxation process in the dielectric spectra, in accord with the experimental results. The resulting relaxation step in $\varepsilon'(\nu)$ reveals a very large value of the limiting low-frequency $\varepsilon'$ (sometimes termed "colossal dielectric constant" [22]. The latter is dominated by the capacitance of the surface layers which is large due to their small thickness. At high frequencies, their capacitance becomes essentially shorted and the intrinsic bulk properties are detected. This effect is also responsible for a crossover of $\sigma'(\nu)$ from small values at low frequencies (dominated by the high layer resistance) to larger, intrinsic values at higher frequencies. Considering the relation $\varepsilon'' \propto \sigma'/\nu$, this causes a relaxation-like peak in the loss spectra. Its spectral shape follows the Debye formula, theoretically predicted for dipolar relaxations, although in fact for such non-intrinsic Maxwell-Wagner relaxations it is not related to dipolar dynamics. In some cases, perfect fits of the experimental spectra cannot be achieved by using a simple RC equivalent circuit to model the blocking electrodes. Introducing a distribution of RC elements or replacing the electrode-RC circuit by a so-called constant-phase element then leads to better fits of the data [17,18,23].

Notably, the above-described equivalent-circuit approach in principle should only provide an approximate description of electrode polarization. In fact, the capacitive layers only form at low frequencies, simply because most ions do not reach the sample boundaries at high frequencies. Nevertheless, the manifestations of electrode polarization in experimental spectra of ionic conductors can be excellently described using this somewhat simplified approach [18,21,24]. Overall, one should be aware that electrode polarization is a non-intrinsic phenomenon and its manifestations in the measured permittivity and conductivity spectra should not be misconceived to reflect intrinsic properties like dipole dynamics. To avoid such misinterpretations, e.g., additional measurements with different sample thicknesses can be helpful [18]. Finally, it should be noted that thin *internal* barriers in biological matter (e.g., cell walls) can cause similar effects, however, usually located at higher frequencies (see discussion of the $\beta$ relaxation in section 2.5).

## 2.4 $\alpha$ Relaxation

In various biological systems like tissues or cell suspensions, the occurrence of a low-frequency process (usually at $\nu < 1$ kHz), termed $\alpha$ relaxation, is reported [1,2,8,12,13,25]. The associated absolute values of the limiting low-frequency dielectric constant are huge (»1000), clearly indicating that it is not due to the reorientation of dipolar entities within the material. A common explanation for the

$\alpha$ relaxation ascribes it to the mobility of "counterions" along the surface of colloidal particles (e.g., cells) within electric double layers formed around them [1,2,12,13,26,27]. In particular, it was pointed out that this relaxation is not due to a simple Maxwell-Wagner effect [26,27]. Various alternative mechanisms for this process were also proposed [2,28]. Notably, due to its occurrence at very low frequencies, usually the $\alpha$ relaxation cannot be detected in dielectric-loss spectra. This is due to the huge contribution of the dc conductivity to $\varepsilon''(\nu)$ at low frequencies ($\varepsilon''_{dc} \propto 1/\nu$, as noted above), preventing the generation of meaningful dc-corrected data by subtraction. Unfortunately, the $\varepsilon'$ spectra in this low-frequency region are often dominated by EP, also hampering the unequivocal observation of any intrinsic low-frequency processes.

## 2.5 $\beta$ Relaxation

Molecules of biological significance often carry a permanent dipole moment and, when dissolved in water, can reorient with a rate determined by temperature and viscosity of the fluid. As discussed in the following section, the water molecule also is characterized by a permanent dipole moment, and, hence, in biological fluids dielectric dispersion phenomena can be expected due to relaxations of the large biological and the considerably smaller water molecule. Indeed, since the early works by Oncley and co-workers [29,30,31], it is well documented that aqueous solutions containing biological molecules reveal at least two dispersion regions in the MHz and GHz frequency regimes, termed $\beta$ and $\gamma$ relaxation, respectively. The $\gamma$ process, reflecting water dynamics, will be discussed in the following section. As schematically depicted in Fig. 1, specifically in protein solutions the $\beta$ process can be attributed to reorientational motions of the whole protein molecule in the viscous environment of the aqueous solution. Since Oncley's pioneering works, many dielectric spectra were collected specifically for aqueous protein solutions (see, e.g., [32,33,34,35]), all of them pointing towards the relaxation of these macromolecules, characterized by a dipole moment of some 100 D and by a relaxational frequency of some MHz. The temperature dependence of the relaxation times, which can be derived from the loss-peak frequencies, in most cases (at least in the rather narrow temperature regimes being accessible in aqueous solutions) follows purely thermally-activated Arrhenius-type of behavior, characterized by energy barriers of ~20 kJ/mol or ~0.2 eV.

Relaxations in a comparable frequency regime are also observed in all kinds of tissue, e.g., liver, fat, muscle, lung, and in blood [1,2,6,12,13,24,36]. In most cases, they are significantly stronger than those in protein solutions and must be explained by Maxwell-Wagner effects, analogous to the electrode polarization treated in section 2.3. In this case, however, the relevant thin layers, giving rise to polarization effects, are thought to be the cellular membranes. They act like large capacitors and are charged through the electrolyte with the charging time varying inversely with the ionic conductivity. Again, the problem is that this type of dispersion, commonly observed in heterogeneous media, looks analogous to Debye-like relaxations of dipolar entities. Schwan pointed out that, in principle, it is possible to determine its microscopic nature from an analysis of the dependence of the dielectric parameters on temperature or cell concentration [37].

Nowadays, it seems quite generally accepted that most $\beta$ relaxations in tissues and cell suspensions like blood must be ascribed to Maxwell-Wagner like effects [1,2,24,36]. Thus, dielectric spectra in this regime in principle should be analyzed in terms of equivalent circuits, just as commonly done for electrode-polarization effects [18,21,22,24]. However, for biological matter such $\beta$ relaxations are more commonly fitted by the functions developed for intrinsic dipolar relaxation processes. Then one should be aware that the relaxation times determined by such fits do not reflect the time scales of any intrinsic microscopic dynamics in the investigated materials. Fortunately, there are various models (see, e.g., [38,39,40,41,42]) enabling conclusions about the intrinsic dielectric properties of the different sample regions, based on the fitting parameters resulting from such an approach.

In principle, as there are at least two different mechanisms that may give rise to $\beta$ relaxations, the corresponding features in the spectra of tissues or cell suspensions can be bimodal [1,2]. However, in most cases the interfacial $\beta$ loss peaks due to Maxwell-Wagner effects are significantly stronger than those caused by dipolar protein dynamics [2,13]. Thus, in tissues and cell suspension, the protein-related $\beta$ process is usually not detected in the spectra. Notably, even without assuming any dipolar reorientations of macromolecules, multiple Maxwell-Wagner-related $\beta$ peaks can arise due to the non-spherical shape of cells such as erythrocytes [42]. For example, in blood (section 3.3) some hints at a bimodal $\beta$ relaxation were reported [24]. In Fig. 1, the possible bimodal nature of the $\beta$ relaxation is taken into account by the schematic indication of a double-peak at low frequencies.

Finally, we want to mention that in some NMR investigations of proteins, conformational selection dynamics was reported, which is slower than the reorientational $\beta$ relaxation and seems to play an important role for molecular recognition [43,44,45]. In Ref. [5], indications for such dynamics were also found by dielectric measurements of a protein solution and termed "sub-$\beta$ relaxation".

## 2.6 $\gamma$ Relaxation

The $\gamma$ relaxation occurs in water-containing biological matter and is observed at frequencies around 20 GHz [1,2,7,8,12,13]. Pure water at room temperature reveals a prominent relaxational feature close to this frequency [46,47,48] (see section 3.1 and Fig. 2). Therefore, it is obvious that the $\gamma$ relaxation of biological matter and the main relaxation of pure water are caused by the same dynamical process. In the biophysics literature, this is commonly assumed to be the reorientational motion of essentially unbound water molecules, which indeed corresponds to the most common interpretation of the 20 GHz relaxation of pure water (see section 3.1 for a brief discussion of alternative explanations).

In various studies of biological materials extending to sufficiently high frequencies, certain deviations of the spectral features of the $\gamma$ relaxation, compared to the main relaxation of pure water, were revealed (see, e.g., [7,24,49,50,51]). In addition to a reduction of the peak amplitude, which may be explained by the trivial substitution effect of water by other constituents in the sample, this also includes the broadening of the loss peak and its occurrence at somewhat different frequencies. By investigating these changes, valuable information concerning the complex interactions of biological and water molecules can be gained, which are of high relevance for the functionality of biological systems [7].

## *2.7 $\delta$ Relaxation*

Between the $\beta$ and $\gamma$ dispersions, the dielectric spectra of some biological materials exhibit another process, which is roughly located in the 100 MHz regime [2,12,13,17]. Its existence was unequivocally revealed only after the discovery of the $\beta$ and $\gamma$ relaxations [52,53,54], and, thus, it was termed $\delta$ relaxation. Often, its signature in the real part, $\varepsilon'(\nu)$, is rather weak and evidenced by a shallow continuous decrease with increasing frequency, while in $\varepsilon''(\nu)$ a small, separate peak or at least a shoulder shows up, as indicated in Fig. 1. In the loss spectra, in most cases the $\delta$-relaxation peak is significantly weaker than the neighboring $\beta$ and $\gamma$ peaks and, in addition, it is often hidden by the conductivity contribution. The $\delta$ process is mostly detected in protein solutions (e.g., Refs. [35,52,53,54,55,56,57,58]; see section 3.2 for an example) but was also reported, e.g., for blood [12,36] (however, the comprehensive broadband spectra of blood reported in Ref. [24] could be explained without invoking a $\delta$ relaxation; see section 3.3).

The most common mechanism considered for the explanation of the $\delta$ relaxation, especially in the earlier literature, is the dynamics of bound water, existing in various types of biological matter [35,52,54,56,59]. For example, proteins are known to exhibit a hydration shell of water molecules bound to the surface of the protein molecule via hydrogen bonds. In the spectra, the $\delta$ relaxation shows up at frequencies below that of the $\gamma$ relaxation, which is reasonable because bound water molecules should move more slowly than free molecules. Over the years, the interpretation of the $\delta$ relaxation in terms of bound water gained increasing complexity, e.g., assuming that bound-water relaxations are supplemented by fluctuating polar side groups [59]. There are also reports of multiple $\delta$ relaxations which were ascribed to shells with different degrees of bonding to the protein [13,33,34,56]. Interestingly, the dynamics of bound water was also discussed (e.g., in Refs. [60,61,62,63,64]) in the context of the glass transition and the controversial fragile-to-strong crossover of supercooled water [65,66], which cannot be directly investigated in bulk water due to rapid crystallization occurring within the so-called no-man's land at about 150–235 K [67].

Especially in recent years, there are hints that $\delta$ relaxations can also arise due to other mechanisms such as internal-protein or protein-side-chain dynamics, which can exist in addition to bound-water relaxations [17,68,69]. Moreover, collective protein-water motions were considered [35,63,69]. It was even suggested that, in aqueous solutions, the dynamics of hydration water cannot be directly detected using dielectric spectroscopy [68].

Overall, it seems clear that multiple contributions can lead to the $\delta$ relaxation which, thus, often is multimodal. However, due to their mutual overlap in the spectra and the superposition with the adjacent $\beta$ and $\gamma$ relaxations, it is impossible to unequivocally resolve all these contributions using dielectric spectroscopy. Thus, additional methods like nuclear magnetic resonance or molecular-dynamics simulations are helpful for definite conclusions on the dynamics of biological matter in the intermediate frequency range between $\beta$ and $\delta$ relaxation [68,70].

## *2.8 Resonances*

Phonon-like or vibrational excitations of intra- and intermolecular origin are characteristic of the short-time dynamics in all types of condensed matter and, hence, also occur in bio-related materials. As schematically indicated in Fig. 1, these Lorentzian-type excitations, with well-defined eigenfrequencies and in most cases low damping, usually occur at frequencies beyond 1 THz, corresponding to ~ 33 cm$^{-1}$ or ~ 4 meV. These excitations can be subdivided into inter- and intramolecular excitations: The former can be characterized as phonon-like translational or librational modes with wave-vector dependent dispersion. Intramolecular excitations are local excitations of stretching ($v$), scissoring ($\delta$), rocking ($\rho$), wagging ($\omega$), or torsional ($\tau$) type, where the former two can have symmetric and antisymmetric character. These local intramolecular excitations, which usually depend mainly on the bonding to next-nearest atomic neighbors, reveal almost no dispersion effects, appear in a frequency regime from 10–100 THz, and very often are well separated in energy from the phonon-like intermolecular excitations. This enormous zoo of excitations mainly is studied using optical (THz, FIR, IR, and Raman) and inelastic neutron-scattering techniques. The crystal symmetry usually determines whether the modes are IR or Raman active.

A three-dimensional molecular crystal with $M$ molecules per unit cell and $N$ atoms per molecule will exhibit $3MN$ degrees of freedom. Clearly, even the spectra of simple molecular crystals reveal an astonishing complexity, and often the microscopic origin of specific excitations can hardly be determined. Even a single molecule with $N$ atoms will show $3N$ - 6 internal modes. This will be discussed in some detail in the analysis of the THz spectra of amino acids (see section 3.4). Amino acids are the building blocks of proteins which consist of hundreds of them, and it seems impossible to realistically model the excitation spectra of proteins on the basis of microscopic details [71,72].

In the light of the following section, here we would like to add a comment on low-lying optical phonons in molecular crystals. Very often intermolecular optical excitations involve displacements of complete molecules with relatively heavy masses and, hence, will be characterized by very low eigenfrequencies. These excitations will significantly contribute to the phonon density of states yielding characteristic and sometimes strong deviations from the expected Debye-type $T^3$ behavior of the specific heat, which should not be confused with the boson peak expected in amorphous solids. These large excess-heat capacities were clearly documented for some amino acids [11]. These excess contributions at low temperatures, resulting from well-defined optical phonons, raise some doubts whether the boson peak reported to occur in some proteins [73] results from the glassiness of the samples or just stems from low-lying phonons, which seem to be generic excitations in bio-related matter.

## *2.9 Boson Peak*

It is well known that dielectric spectra of disordered matter extending up to THz frequencies reveal a universal additional loss-peak around 1 THz. In the literature on supercooled liquids, it is often termed boson peak [14,74,75,76,77], analogous to the corresponding feature observed in the scattering function and susceptibility measured by neutron and light scattering [78,79]. Various competing explanations of its microscopic origin were proposed (e.g., Refs. [80,81,82,83,84,85]). In the frequency range of the boson peak, *crystalline* matter usually reveals vibrational excitations like phonons, showing up by rather narrow Lorentz peaks in the loss spectra. While the boson peak is much broader than such typical resonance excitations, most of the proposed explanations for its occurrence assume a phonon- or vibration-related origin.

The boson peak is mostly investigated for supercooled liquids, i.e., at temperatures below the melting point. There, the structural relaxation process (termed $\alpha$ relaxation in glass-physics and usually closely coupled with the reorientational motions of the molecules in dipolar liquids) is located at low frequencies, not obscuring the boson peak. Recently, the latter was proposed to even show up in all kinds of liquids, including water around room temperature [48]. However, in the dielectric spectra, water's boson peak is partly superimposed by its very strong main relaxation peak (usually ascribed to its dipolar relaxation, see section 3.1) which, at room temperature, is located at very high frequencies (~20 GHz).

As discussed in section 2.6, biological systems containing water exhibit essentially the same process around 20 GHz as pure water, which is termed $\gamma$ relaxation in biophysics. Thus, it is reasonable that their dielectric response also resembles that of water concerning the boson peak (section 3.1), which should give rise to an additional contribution at the right flank of the $\gamma$ relaxation. Indeed, data extending to sufficiently high frequencies seem to be consistent with such a scenario [50,86], and the corresponding loss peak is schematically indicated in Fig. 1 (denoted "BP"). The occurrence of the boson peak in biological matter is also corroborated by its recent unobscured detection in dry protein powder using THz spectroscopy [87]. A boson peak was also found by scattering experiments for proteins and other biomolecules [70] (however, see the critical remarks concerning the boson peak in biological matter in the preceding section).

## 3 Experimental Examples

### *3.1 Pure Water*

More than half of the body mass of most animals is made up of water [88]. Thus, to understand the properties of biological matter in organisms, one has to understand its interactions with water. As mentioned in section 2.6 and recently pointed out in Ref. [7], information about such interactions can be obtained, e.g., by a comparison of the main relaxation process, revealed in the dielectric spectra of pure water, with the corresponding dynamics in biosystems, the $\gamma$ relaxation. Therefore, knowledge of water's dielectric spectra is essential. This is also valid for other fields, e.g., the everyday application of microwave cooking or the absorption of radar waves by clouds [89,90,91].

Figure 2 shows dielectric-loss spectra of pure water covering frequencies from 100 MHz to 20 THz and measured at various temperatures [48]. Data obtained using different experimental setups were combined to arrive at these spectra. In contrast to most other broadband data in literature, results for different temperatures are provided, which were collected by a single workgroup using the same sample preparation. The main feature of the spectra of Fig. 2 is the well-known prominent loss peak of water, which at room temperature is located at about 20 GHz. As mentioned in section 2.6, it is observed within the same frequency range as the ubiquitous $\gamma$ relaxation of water-containing biological matter, clearly pointing to the same principle microscopic origin. Nevertheless, amplitude, peak frequency, and width of the $\gamma$ relaxation can be somewhat modified compared to pure water due to interactions with ions, macromolecules, or cellular structures [7]. The temperature-induced shift of the peak revealed in Fig. 2 is typical for relaxational processes and essentially mirrors the slowing down of the molecular dynamics due to reduced thermal activation at low temperatures [14,15].

As mentioned in section 2.6, in the biophysical literature, the $\gamma$ dispersion is commonly ascribed to the dipole dynamics of unbound water molecules [1,2,13,56] as also existing in pure water. Nevertheless, one should be aware that part of the current literature on pure water also considers other interpretations of its main relaxation process (e.g., Refs. [92,93,94,95,96,97,98,99]). The most prominent one is analogous to the interpretation of the dominant Debye relaxation detected in many monohydroxy alcohols [100,101,102]: It is believed to mirror the motions of dipolar clusters formed by hydrogen-bonded molecules [103], while the structural relaxation from essentially single-molecule motions (termed "$\alpha$ relaxation" outside of biophysics) shows up as a minor relaxation peak at higher frequencies only. However, for pure water such an interpretation is disputed and arguments against this scenario were provided, e.g., in Refs. [48,104].

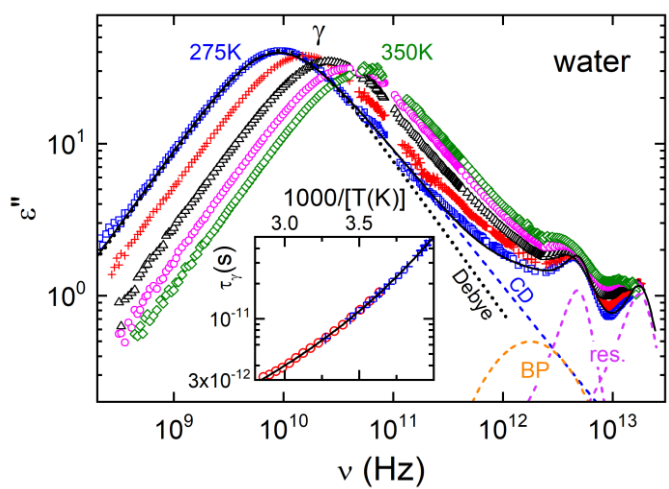

**Fig. 2** Dielectric-loss spectra of pure liquid water as measured at various temperatures (from left to right: 275, 290, 310, 330, and 350 K) (data from Ref. [48]). The dotted line shows a fit of the main peak at 275 K by the Debye function, as often assumed for pure water, which, however, widely fails to describe the right peak flank. For the complete spectrum at 275 K, the solid line shows a fit with the sum of a CD function (for the main peak), a log-normal peak function (for the boson peak), and two Lorentz functions (for the resonances at ~5 and 19 THz) [48]. The dashed lines depict the different contributions to the overall fit curve. The inset presents the average relaxation times derived from the CD fit parameters (circles) (data from Ref. [67]) and additional $\tau(T)$ data from Ref. [117], extending into the supercooled region. The line represents a fit with the VFT function [67]. Please note that the main peak of water is named "$\gamma$" in the biophysical literature while it is denoted $\alpha$ or Debye peak in other fields.

Traditionally, the 20 GHz relaxation of water is fitted by a Debye function derived within the Debye theory of dipolar relaxation [105]. However, newer measurements with sufficient precision at frequencies up to the THz range revealed deviations of the experimental data from the Debye function at the high-frequency flank of the 20 GHz loss peak [92,93,95,106]. They cannot be explained by contributions of the known vibration-related resonances observed beyond THz, usually described by Lorentz functions. This becomes especially clear for spectra close to the freezing point of water as provided in Ref. [48] (see 275 K curve in Fig. 2) where the main peak has shifted to lower frequencies while the resonances remain largely unaffected by temperature. As revealed by the dotted line in Fig. 1, at the right flank of the main loss peak, already for frequencies that are about four times larger than the peak frequency, deviations from the Debye fit start to show up. It should be noted that there is no fundamental reason to assume the validity of the Debye model for pure water. This becomes obvious when considering that for nearly all other dipolar liquids, deviations from Debye behavior, especially pronounced at the high-frequency flanks of their main relaxation peaks, are very common [14,15,107]. It is nowadays quite widely accepted that the corresponding broadening of the relaxation peaks, exceeding the half width of 1.14 decades predicted by the Debye model, is due to a distribution of relaxation times caused by the disordered structure of liquids [108,109]. This should apply to water just as to other dipolar liquids. Indeed, the water spectra can be significantly better fitted by the empirical Cole-Davidson (CD) function [110], often used for dipolar liquids [14,15,111] (see blue dashed line denoted "CD" in Fig. 2). It matches, e.g., the 275K spectrum for frequencies up to about 1.5 decades beyond the peak frequency.

Anyway, Fig. 1 demonstrates that, even when using a CD function for the main peak (blue dashed line), deviations from the fits at high frequencies still show up, due to significant excess intensity in $\varepsilon''(\nu)$ at ~1 THz. In literature, the main loss feature of water, including this excess contribution, was mostly fitted by the sum of two Debye functions [92,112,113], probably sticking to the Debye shape of the main peak due to historical reasons. In general, broadband spectra of water were described by many different combinations of various relaxation and resonance contributions, and good fits of the experimental data were achieved with partly very different approaches [93,95,96,97,98,99,106,113]. Of course, one should be aware that, by combining a sufficiently large number of peak functions such as Debye or Lorentz, almost any experimental curve can be fitted.

In Ref. [48], we have pointed out the similarity of the water dielectric spectra to those of many glass forming dipolar liquids [14,15,74,111,114,115] (e.g., glycerol at 413 K [111]) if the latter are heated to temperatures where their main loss peaks are located in the GHz region (cf. Fig. 3a). These glass formers can be easily supercooled below their melting point $T_m$. As schematically indicated in Fig. 3b, this supercooling shifts their main relaxation peak to low frequencies, which reveals the origin of the excess intensity around 1 THz: It is due to a separate, essentially temperature-independent loss peak, clearly observed in their low-temperature spectra just in this frequency region. It was termed the boson peak [74] in analogy to the corresponding feature found by scattering methods [78,79] (see section 2.9). Its microscopic origin is still controversial. Unfortunately, pure water cannot be easily supercooled to relieve its suggested boson peak from the superposition of the main relaxation at 20 GHz [67]. However, supercooling becomes possible when adding certain amounts of salt to water [48,67,116]. At room temperature, the dielectric-loss spectra of such solutions exhibit just the same excess contribution as pure water at the right flank of their main peak at ~20 GHz [48]. When they are supercooled, indeed this shoulder develops into a peak at about 1 THz, which can be interpreted as the boson peak of water [48]. As mentioned in section 2.9, it seems reasonable that this contribution is also present in biological matter containing water (cf. Fig. 1).

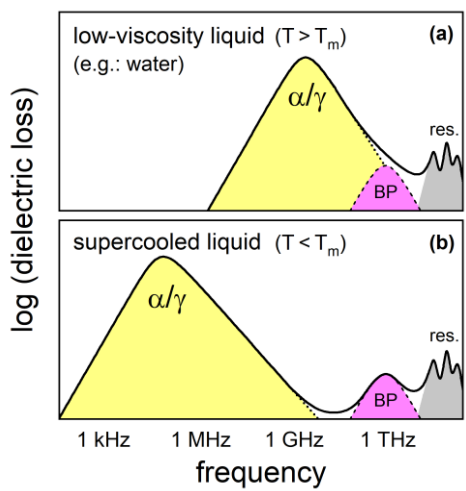

**Fig. 3** Schematic dielectric-loss spectra of a low-viscosity dipolar liquid (**a**) (e.g., water around room temperature [48] or glycerol at $T \sim 410$ K [111]) and of a supercooled liquid (**b**) (e.g., water with LiCl at ~170 K [48] or glycerol at ~240 K [111]). The main peak is named $\alpha$ relaxation in the field of supercooled liquids and $\gamma$ relaxation in biophysics for materials containing unbound water. "BP" stands for the boson peak and "res." denotes high-frequency resonances. The fast process of supercooled liquids, discussed in glass physics to appear in the minimum region between $\alpha$ and boson peak [14,15], is not indicated in the figure.

To illustrate the suggested approach for pure water, the solid line in Fig. 2 is a fit of $\varepsilon''(\nu)$ at 275 K with the sum of four contributions [48]: A CD function to cover the main relaxation, two Lorentz functions for the vibrational infrared resonances, and a log-normal peak function for the boson peak. The latter is entirely phenomenological and is able to account for the fact that the boson peak in dipolar liquids is narrower than a Debye peak but broader than a resonance peak [14,15,74,111,114,115].

From fits of the loss peaks in Fig. 2, the temperature-dependent relaxation times of the main dynamic process of pure water can be deduced. The inset of Fig. 2 shows an Arrhenius plot of the resulting $\tau(T)$, also based on spectra collected at further temperatures [48,67] (circles). The plusses represent additional data reported in Ref. [117], extending the temperature range into the supercooled region, down to 252 K, below which the mentioned no-man's land starts [67]. The curved shape of this combined data set provides clear evidence for significant deviations from Arrhenius temperature dependence, $\tau(T) \sim \exp(E/T)$. Here $E$ is an energy barrier (in K) that has to be overcome by thermal activation for the corresponding motion to happen. Assuming that the main relaxation of water is due to reorientations of the water molecules, $E$ should be largely determined by the energy needed to break the intermolecular hydrogen bonds. However, the found deviations from Arrhenius behavior imply that the situation in water is more complex. Interestingly, non-Arrhenius behavior as found for $\tau(T)$ of pure water is also a very common property of other dipolar liquids where it was mainly investigated in the supercooled state but also exists above $T_\mathrm{m}$ [14,15]. It is often ascribed to an increase of the cooperativity of molecular motions with decreasing temperature [118,119], which also seems to apply to water.

The line in the inset of Fig. 2 is a fit with the empirical Vogel-Fulcher-Tammann (VFT) function, $\tau(T) \sim \exp[DT_\mathrm{VF}/(T-T_\mathrm{VF})]$, often applied to dipolar liquids [14,15, 111,114,118,120,121]. Here $D$ is termed strength parameter and $T_\mathrm{VF}$ is the Vogel-Fulcher temperature. From $D$ the so-called fragility index of a liquid can be calculated [120], which is a measure of the degree of deviations from Arrhenius temperature dependence of $\tau(T)$. We obtain a high value of $m = 175$ [67], making liquid water a very "fragile" liquid within the often-employed classification scheme introduced by Angell [122], in accord with earlier reports [65]. However, one should be aware that the fragility of pure water and the closely related fragile-to-strong crossover suggested in its supercooled state are highly controversial topics [65,66,123,124,125,126,127] (see Ref. [67] for a more detailed discussion). Notably, one approach to clarify the open questions about the behavior of pure water in the no-man's land and its controversial glass transition is the investigation of the dynamics of water within biological matter which often can be easily supercooled (e.g., [60,61,62,63,128]). However, a detailed discussion of such studies is out of scope of the present work.

## *3.2 Broadband Spectra of a Protein Solution*

Proteins are major components of all cells and play crucial roles in almost every biological process. Via dielectric spectroscopy, they are studied not only in the pure, solid state (mostly as powder), but also dissolved in water, which more closely resembles in vivo conditions. This is essential because water is known to strongly affect the biological functions of proteins, although the details of these interactions are only partly understood [7,34,64,129,130]. Therefore, investigations of dissolved and hydrated proteins and of protein-water interactions are very active research fields.

Figure 4 shows an example for permittivity and conductivity spectra of a typical aqueous protein solution (5 mmol/l lysozyme) at body temperature, covering a very broad frequency range [86]. Such broadband spectra are essential for the investigation of the many different aspects of protein dynamics, e.g., dipolar relaxation processes of the protein molecules themselves and of bound and unbound water molecules (cf. Fig. 1). At low frequencies, the dielectric constant $\varepsilon'(\nu)$ approaches unrealistically large values beyond $10^8$ (Fig. 4a), clearly indicating a non-intrinsic origin. In fact, electrode polarization, caused by the blocking of translational ionic motions at the metallic electrodes, generates this huge increase (see section 2.3). A detailed discussion of this effect, as detected in a similar lysozyme solution and various other materials, in terms of a equivalent-circuit analysis was provided in Ref. [18]. Moreover, the decrease of the conductivity below about 1 kHz, revealed in Fig. 4c, is also due to electrode polarization and simply caused by the concomitant reduction of the time-averaged ionic mobility at low frequencies. Above 1 kHz, $\sigma'(\nu)$ remains approximately constant up to about 1 GHz, signifying the ionic dc conductivity of this solution. Due to the general relation $\varepsilon'' = \sigma'/(2\pi \nu \varepsilon_0)$, mentioned in section 2.2, the dc charge transport leads to $1/\nu$ behavior in the loss spectra (evidenced by the linear decrease between about 1 kHz and 10 GHz in Fig. 4b) while the electrode polarization causes the crossover to weaker $\varepsilon''(\nu)$ variation seen at lower frequencies, $\nu < 1$ kHz.

The steplike decrease in $\varepsilon'(\nu)$ and the peak in $\varepsilon''(\nu)$, revealed at the highest frequencies in Figs. 4a and b, are typical signatures of a relaxation process. Due to the mentioned direct relation of $\varepsilon''$ and $\sigma'$, the loss peak corresponds to a shoulder in the conductivity spectra (Fig. 4c). This process can be identified with the $\gamma$ relaxation of this protein solution [56,57], commonly assumed to arise from reorientational motions of unbound water molecules (but see section 2.6 for a possible alternative scenario discussed for pure water). This assignment becomes immediately obvious from the close agreement of the loss-peak frequencies and amplitudes in Fig. 4b with those of pure water (Fig. 2). This is illustrated in the inset of Fig. 4 showing $\varepsilon''(\nu)$ of the 5 mmol lysozyme solution (open circles) and of pure water (line) in the $\gamma$-peak region which agree very well. As discussed in Ref. [57], in fact the peak amplitude of the protein solution is a few percent smaller than that of pure water (not visible in the inset of Fig. 4, covering more than one decade in $\varepsilon''$), which is due to its slightly reduced water concentration.

At first glance, based on Fig. 4, only three dynamic processes seem to exist in this protein solution, none of them providing direct information on the protein itself: the electrode polarization which is non-intrinsic, the $\gamma$ relaxation, mirroring motions of free water molecules just as in pure water, and the dc conductivity which is due to translational motions of dissolved mobile ions stemming from residual salt contents. However, as noted in sections 2.1 and 2.2, subtracting the dc contribution in the loss spectra can uncover additional processes (see Refs. [5,57] for possible caveats concerning the dc subtraction). The inset of Fig. 4 shows such corrected data for the 5 mmol solution (closed circles) and for a 3 mmol solution (plusses). The dc subtraction exposes additional data points lying at the left flank of the $\gamma$ peak. Deviations from this flank revealed below about 1 GHz indeed point to at least one additional process at lower frequencies. In fact, these deviations are due to the $\delta$ relaxation as will be discussed below.

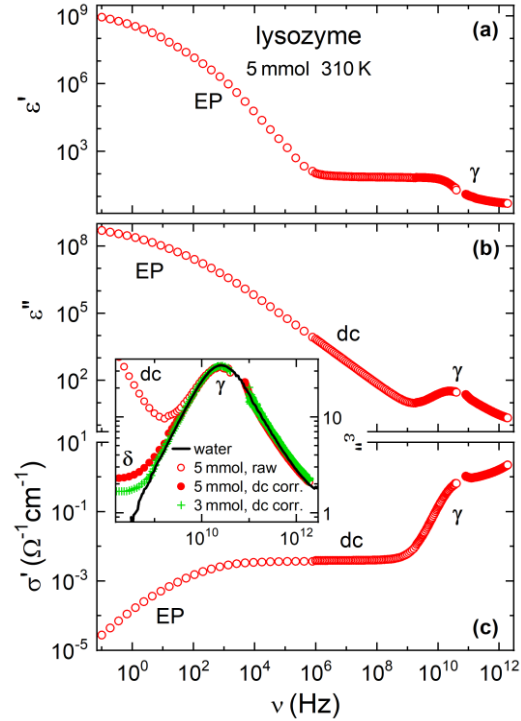

**Fig. 4** Broadband dielectric spectra of a 5 mmol/l lysozyme aqueous solution, showing (**a**) the dielectric constant (real part of the permittivity), (**b**) the dielectric loss, and (**c**) the real part of the conductivity as measured at 310 K (data from Ref. [86]). The contributions from electrode polarization, $\gamma$ relaxation, and dc conductivity (the latter only existing in $\varepsilon''$ and $\sigma'$) are denoted in the figure. The inset provides a magnified view of the $\gamma$-relaxation region (open circles). The closed circles represent the same data set, corrected for the dc conductivity, and the plusses indicate dc-corrected data for a 3 mmol solution at the same temperature. Both corrected curves reveal the onset of the $\delta$ relaxation (cf. Fig. 5) at low frequencies. In addition, the line in the inset shows $\varepsilon''(\nu)$ of pure water at 310 K [48].

Figure 5a shows $\varepsilon'(\nu)$ of a 3 mmol lysozyme solution above 1 MHz, where electrode polarization can be widely neglected [57,86]. Data for three temperatures are provided, revealing the typical shift of the $\gamma$-relaxation step, observed around 10 GHz, to lower frequencies upon cooling, as expected for thermal activation. In Fig. 5b, we show the corresponding loss spectra, corrected for the dc-conductivity contribution by subtracting $\varepsilon''_{dc} \propto \sigma_{dc}/\nu$ (see section 2.2). In addition to the $\gamma$ peak, which also exhibits the temperature-induced shift seen in $\varepsilon'$, two additional relaxations at lower frequencies are revealed in these dc-corrected data. Most obvious is a peak occurring around 10 MHz. It can be ascribed to the $\beta$ process of this protein solution [57]. As mentioned in section 2.5, $\beta$ relaxations typically show up just in this frequency region and are mostly considered to arise from two possible mechanisms: Maxwell-Wagner relaxations due to internal heterogeneities, e.g., cell membranes, and/or reorientational motions of large dipolar molecules [1,2,6,13,35]. Obviously, the latter applies to the present solutions containing dipolar protein macromolecules. The shift of the $\beta$-relaxation peaks with temperature, seen in Fig. 5b, again indicates that the protein dynamics is essentially thermally activated.

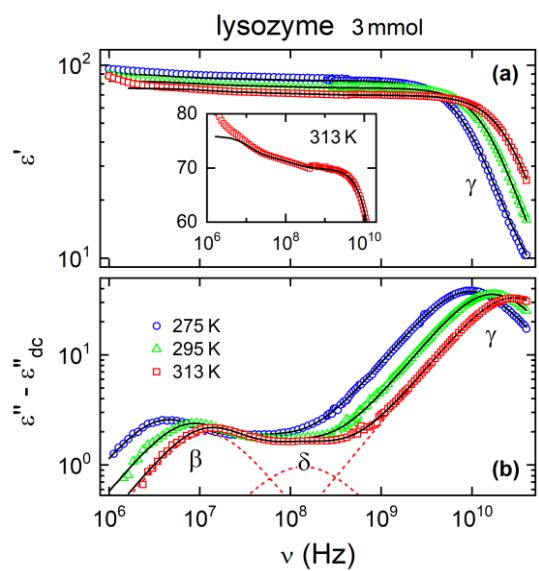

**Fig. 5** Dielectric spectra of a 3 mmol/l lysozyme aqueous solution, showing (**a**) the dielectric constant and (**b**) the dc-corrected dielectric loss, as measured at three temperatures (data from Refs. [57,86]). The solid lines are fits with the sum of a Debye and two CC functions for the $\beta$, $\delta$, and $\gamma$ relaxations, respectively. They were simultaneously performed for the real and imaginary part. For 313 K, the dashed lines in (b) show the contributions of the three processes to the overall fit. The inset presents a magnified view of $\varepsilon'(\nu)$ at 313 K in the region of the low-frequency plateau.

Between the two well-pronounced $\beta$ and $\gamma$ peaks, the loss spectra in Fig. 5b reveal significant additional intensity that is due to the $\delta$ relaxation [57]. As discussed in section 2.7, the origin of the $\delta$ process is controversial, and different processes may contribute in this frequency region. Thus, in principle it could be multimodal, e.g., composed by a sum of several Debye contributions as assumed in some publications [33,35,56,58,69]. In fact, an investigation of a frozen lysozyme solution, where the ionic conductivity is strongly reduced, revealed unequivocal evidence for its bimodal nature [63]. A comparison of those results with data for liquid lysozyme solutions [57] and hydrated powders [86] provided evidence that the faster component of the $\delta$ relaxation arises from the dynamics of loosely bound hydration water [63]. Moreover, in Ref. [63] collective protein-water motions [35,69,131] were suggested to be the origin of the slower component. In the spectra for liquid lysozyme solutions shown in Fig. 5, the superposition of the $\delta$ relaxation by the $\beta$ and $\gamma$ processes obviously is too strong to enable conclusions about its possible bimodal nature (see also discussion of the performed fits given below).

The loss-peak amplitudes of the $\beta$ and $\delta$ processes are more than one decade smaller than for the $\gamma$ relaxation. Therefore, corresponding relaxation steps are not immediately evident in $\varepsilon'(\nu)$ (Fig. 5a), which, between about 1 MHz and 3 GHz, is dominated by the large static dielectric constant $\varepsilon_s \approx 70$–$80$ of the $\gamma$ process, mirroring the large $\varepsilon_s$ of pure water [48]. Nevertheless, the inset of Fig. 5a, showing a magnified view of this region for 313 K, reveals faint indications for a sigmoidal decay of $\varepsilon'(\nu)$ around 10 MHz, which is due to the $\beta$ process. At lower frequencies, it is partly superimposed by the onset of electrode polarization and at higher frequencies, the weak $\delta$ dispersion shows up, followed by the onset of the much stronger $\gamma$-relaxation step. (The anomaly close to 500 MHz is due to a slight mismatch of about 1 % between the results from the employed low and high-frequency devices; see Refs. [57,86] for experimental details.) Thus, from the $\varepsilon'$ spectra alone, it would be impossible to draw unequivocal conclusions about the existence of the $\beta$ and $\delta$ relaxations in this protein solution. This does not hold for protein samples where the ionic conductivity is less dominant, e.g., for solutions measured below the freezing point of water as shown in Ref. [63].

The solid lines in Fig. 5 are fits with the sum of one Debye and two Cole-Cole (CC) [132] functions, accounting for the $\beta$, $\delta$, and $\gamma$ relaxations, respectively. The empirical CC function gives rise to symmetrically broadened loss peaks and is often employed for disordered matter [15,132,133]. As mentioned in section 3.1, such non-Debye behavior is assumed to be caused by different environments sensed by each molecule, giving rise to a distribution of relaxation times [108,109]. Concerning the $\beta$ process, which is ascribed to protein reorientations, concentrations of few mmol/l imply such small amounts of dissolved protein molecules that direct intermolecular protein interactions appear unlikely. Thus, each macromolecule basically "sees" the same time-averaged environment of very rapidly (related to the $\beta$-relaxation time) fluctuating water molecules [57]. This rationalizes the use of the Debye function for the description of the $\beta$ peaks, in agreement with previous publications [33,35,50,56,58]. As noted in Ref. [57], the $\gamma$ relaxation in Fig. 5 can be best accounted for by a CC function, in accord with later reports on different protein solutions [51,68]. Notably, just as for pure water (section 3.1), in some publications multiple Debye peaks were used to fit dielectric spectra of protein solutions in the $\gamma$-relaxation region (e.g., [50,58]). However, when considering Occam's razor, we think a single CC function, assuming only one instead of two dynamic processes and involving fewer fit parameters, is preferable. For the $\delta$ process, a CC function was employed in the fits, too. As mentioned above, in fact it may be multimodal, e.g., composed by a sum of several Debye contributions as assumed in some publications [35,56,58]. Nevertheless, the $\delta$ relaxation in the spectra of Fig. 5 can be well modelled by a single CC function, and using multiple peaks would lead to parameters with limited significance in the present case.

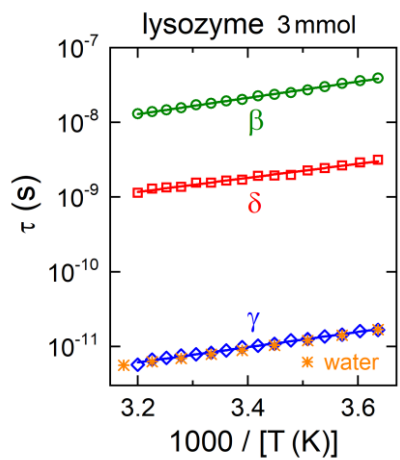

**Fig. 6** Temperature-dependent relaxation times of the three processes in the 3 mmol lysozyme solution as deduced from the dielectric spectra (open symbols; data from Ref. [57]). The lines are Arrhenius fits with energy barriers 0.22, 0.19, and 0.21 eV for the $\beta$, $\delta$, and $\gamma$ processes, respectively. The stars show data for pure water [67].

The temperature dependences of the relaxation times obtained from the fits of the dielectric spectra are shown in Fig. 6 (open symbols), using an Arrhenius representation, $\log_{10}(\tau)$ vs. 1000/T. The found linear behavior indicates thermally activated behavior. However, one should be aware that the temperature range accessible in these aqueous liquids is quite limited (compared, e.g., to glass forming liquids which can be supercooled [14,15]), which is why non-Arrhenius behavior in a broader temperature range cannot be excluded. The stars in Fig. 6 are data for pure water [67]. They agree well with $\tau(T)$ of the $\gamma$ process, again evidencing its origin due to unbound water-molecule dynamics. Interestingly, the similar slopes in Fig. 6 point to similar hindering barriers for the three relaxational processes in this protein solution (between 0.17 and 0.22 eV). As discussed in detail in Ref. [57], the $\beta$ and $\gamma$ relaxations have almost identical barriers and, thus, are most closely coupled, as expected for rotations of molecules in a solution. In contrast, the $\delta$ dynamics seems to be less governed by that of the solvent molecules and appears to be influenced by interactions with the protein molecules. A more detailed discussion of the relaxation times and various other parameters deduced from the dielectric spectra of different lysozyme solutions is provided in Ref. [57]. For example, the results enable conclusions about the hydrodynamic radius and dipole moment of the protein molecules. Comparing the obtained dc conductivity and the $\gamma$ relaxation time reveals that the ionic conductivity is decoupled from the solute-molecule dynamics, implying a breakdown of the Debye Stokes-Einstein relation [134,135].

Overall, the results exemplarily discussed in the present section demonstrate that broadband dielectric spectroscopy, extending well into the GHz range, can provide valuable information about the complex dynamics in aqueous protein solutions, helping to uncover their microscopic origins.

## *3.3 Broadband Spectra of Blood*

Needless to say, blood is one of the most important liquids ("Blut ist ein ganz besonderer Saft", J. W. von Goethe, Faust I), and in living organisms it serves as a highly functional body fluid. It delivers oxygen to the vital parts of the body, transports nutrients, vitamins, and metabolites, and it also acts as a fundamental part of the immune system. Therefore, the precise knowledge of its constituents and their biological, physical, and chemical properties including the involved dynamic processes are of great importance. On average, approximately 8% of the body weight is blood, consisting mainly of blood plasma (55%) and red blood cells (RBCs) (45%), with an additional small amount of white blood cells and platelets. RBCs or erythrocytes have a biconcave shape with a cell membrane composed of proteins and lipids, filled with erythrocyte cytoplasm with high hemoglobin concentration and without a nucleus. Naively, one could say that RBCs are bags filled with hemoglobin. Blood plasma consists of 90% water with solutes, like proteins, lipids or salts. In modeling the dielectric response of cell suspensions including blood, usually heterogeneous media are assumed: The RBCs are approximated by ellipsoids, with well-defined inner radius, filled with medium, encapsuled by a shell of given thickness and embedded in a homogeneous solution. Cell interior, shell, and solution are assumed to have well-defined dielectric parameters [39].

The dielectric parameters of blood, which have been investigated since more than 100 years, are of relevance for various biotechnological and medical applications, e.g., for the separation of infected from normal blood cells [136], cancer diagnosis [137], for glucose monitoring of diabetes patients [138], for the examination of possible deterioration of preserved blood [139], to determine the sedimentation rate of erythrocytes [140], and to estimate the precise dielectric properties of blood as prerequisite for fixing limiting values for electromagnetic pollution [141], to name just a view. The detailed investigation of the analysis of dielectric properties of blood started with the works of Höber [142] and Fricke [143] measuring capacitance and conductance of cell suspensions at kHz to MHz frequencies. Shortly after World War II, with microwave technology being developed to a high standard, experiments on blood were extended well into the GHz regime [144,145,146,147], allowing to study the dynamics of cell suspensions, including blood at high frequencies. Since then, detailed dielectric work was performed, partly with the emphasis on determining dielectric material parameters of the blood constituents, as the capacitance of the cell membranes and the conductivities of plasma and cytoplasm, but partly also to establish number and microscopic character of the various relaxation processes detected by broadband dielectric spectroscopy. Informative surveys and detailed references to published dielectric work on proteins, tissues, and specifically on blood can be found in [1,2,12,17,24,28,138,148,149,150,151,152].

Broadband dielectric spectra on blood, important for a critical analysis of the involved relaxation processes, were published by Gabriel *et al*. [149] and Wolf *et al*. [24]. The former, however, represent a collection of results from various groups, using different samples, partly with relatively large scatter, and at a single temperature only. In contrast, Wolf *et al*. [24] have investigated blood using identical samples for all frequency ranges, measured as function of temperature and hematocrit value. Results on whole blood collected within the framework of that study are shown in Fig. 7, covering more than 10 decades in frequency and including all important relaxation regimes. The three frames display the frequency dependence of dielectric constant (a), dielectric loss (b), and real part of the conductivity (c) of whole blood on double logarithmic scales at two temperatures (280 and 330 K), as well as the spectra of blood plasma at 280 K. Whole blood clearly exhibits three major dispersion regions as often observed in raw dielectric spectra of all kinds of tissues and cell suspensions [1,2,12,13,28,153,154]. Wolf and coworkers [24] assigned these relaxations with increasing frequency to electrode polarization, $\beta$, and $\gamma$ relaxation. Qualitatively similar results were also obtained for lower RBC concentrations, i.e., smaller hematocrit values [24]. Figure 7 reveals that blood plasma only exhibits two relaxation steps, namely EP and $\gamma$ process, with the $\beta$ process being completely absent [24].

The dipolar strength and temperature dependence of the $\gamma$ dispersion in blood, as shown in Fig. 7, resembles the relaxational dynamics of water molecules as observed in pure water with a maximum of the loss peak close to 20 GHz (section 3.1) [48]. As discussed in section 2.6, this experimental result is similar to most of the high-frequency observations in biological matter in solution. According to Wolf *et al*. [24] the $\gamma$ relaxation of blood can be well described by the symmetrically broadened CC function close to pure mono-dispersive Debye behavior, with a slightly increasing width upon cooling. The temperature dependence of the relaxation time, inversely proportional to the loss-peak frequency, can be well fitted by thermally activated Arrhenius behavior with a hindering barrier of about 0.2 eV. Figure 7 documents that blood plasma, lacking any cells, at 280 K reveals a $\gamma$ relaxation that is of similar strength and peak frequency as found for blood at the same temperature. This further corroborates the assignment of the $\gamma$ relaxation to the dynamics of unbound water molecules (section 3.1).

On decreasing frequency, in the loss spectra (Fig. 7b) the $\gamma$ relaxation exhibits a smooth transition into $1/\nu$ behavior due to dc conductivity (see section 2.2), signified by a dc plateau in $\sigma'(\nu)$ around 100 MHz (Fig. 7c). In blood, this dc plateau depends significantly on temperature as routinely observed in ionic conductors. In blood plasma, this dc plateau extends until the onset of the electrode-polarization contribution at low frequencies, while in whole blood it is limited by the onset of the $\beta$ relaxation (section 2.5). Interestingly, as seen for 280 K in Fig. 7c, $\sigma_{dc}$ of the plasma is of similar magnitude as for blood.

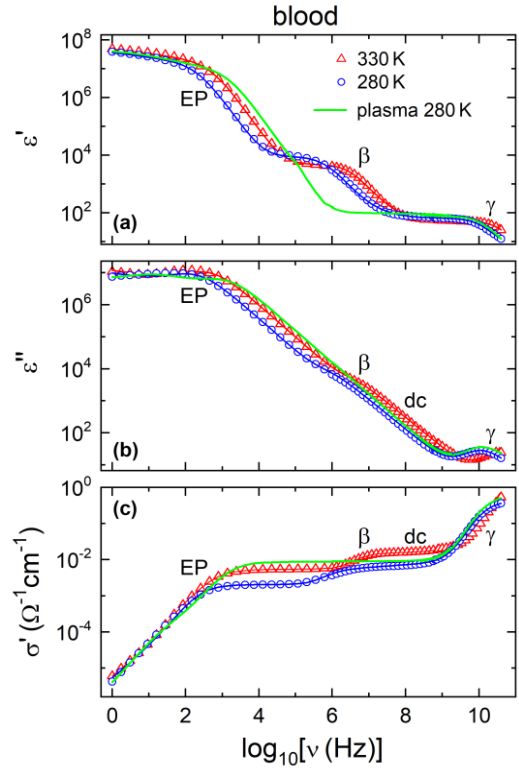

**Fig. 7** Frequency dependence of dielectric constant (a), dielectric loss (b), and real part of the conductivity (c) of whole blood at two different temperatures and of blood plasma at 280 K (the latter from Ref. [24]). The three well-separated dispersion regimes are indicated as electron polarization (EP), and as $\beta$ and $\gamma$ relaxations. The lines are fits (simultaneously performed for $\varepsilon'$ and $\varepsilon''$) assuming a distributed RC equivalent circuit to account for the electrodes [18] and two Cole–Cole functions for the $\beta$- and $\gamma$-relaxations. For the plasma data, only a single Cole-Cole function was used.

This additional dispersion in whole blood, ascribed to a Maxwell-Wagner effect governed by the cell membranes [12,24,36], shows up at MHz frequencies (Fig. 7). In $\sigma'(\nu)$ (Fig. 7c), it is limited by two regions of frequency-independent conductivity. The intrinsic dc conductivity of the sample seen at high frequencies is governed by ionic charge transport within the plasma and cytoplasm As discussed in Ref. [24], the lower conductivity plateau found at frequencies below the $\beta$ dispersion (at about 1 kHz–1 MHz, depending on temperature) is affected by the presence of RBCs in blood, in particular their membranes, locally blocking the charge transport. Typical for many Maxwell-Wagner effects (sections 2.3 and 2.5) [18,21,22], at high frequencies, the capacitive interfaces (here: the cell membranes) become effectively short-circuited and the higher, intrinsic dc conductivity value is detected. The latter is in fact a mixture of the conductivity of the plasma outside of the cells and of the cytoplasm within the cells, the latter having somewhat lower conductivity [24]. This scenario is well supported by the mentioned approximate agreement of this high-frequency plateau value with the dc conductivity of the pure plasma [24] as demonstrated in Fig. 7c for 280 K (the remaining small deviations were suggested to arise from the volume fraction of blood's cytoplasm [24]).

The complete absence of the $\beta$-dispersion effects in the plasma sample (Fig. 7) is also well consistent with Maxwell-Wagner-like effects. However, this could also be said for possible intrinsic relaxations due to dipolar reorientation dynamics of macromolecules (e.g., proteins within the RBCs), which also can lead to $\beta$ relaxations in biological matter (see sections 2.5 and 3.2). Dispersion effects due to these two very different phenomena in principle look similar. However, in the present case, the very high static dielectric constant of blood's $\beta$ relaxation of order $10^4$ revealed in Fig. 7a again strongly speaks in favor of the discussed Maxwell-Wagner-related origin due to cell membranes. One should further note that, for a reorientational relaxation, the conductivity plateau above the corresponding dispersion in $\sigma'(\nu)$ is not at all related to the dc conductivity of the sample, simply because it arises from the right flank of a dipolar relaxation peak in the loss. Thus, within such a scenario it could not be explained why this plateau value in blood approximately agrees with that in the plasma sample (Fig. 7c)

In Ref. [24], the frequency regime of the $\beta$ relaxation of blood was phenomenologically analyzed by a CC function, with parameters that come very close to the mono-dispersive Debye case. It was found that the width parameter and the dipolar strength are almost temperature independent. The relaxation time was revealed to follow Arrhenius-type temperature dependence with energy barriers between 0.11 and 0.15 eV of similar order as for the determined dc conductivity. This agreement is expected for Maxwell-Wagner relaxations, for which the intrinsic conductivity should govern the temperature dependence of $\tau$ [18]. Wolf *et al*. [24] analyzed the detected dispersion effects utilizing the Pauly-Schwan model [39] and obtained reasonable values for the conductivity inside and outside of the cells and for the capacitance of the cell membranes. Interestingly, in Ref. [24] some hints were found indicating that the $\beta$ relaxation of blood is composed of two different, closely overlapping relaxation processes. Such a bimodal $\beta$ relaxation may be explained by Maxwell-Wagner-related models accounting for the non-spherical geometry of the RBCs [42].

Finally, one should note that there are contradicting reports about the presence of an $\alpha$ and a $\delta$ relaxation in blood and blood-related solutions [2,12,36,149,151,155,156,157]. Concerning the $\delta$ relaxation, its presence in blood seems possible because it contains proteins, which should be surrounded by bound water and which may also reveal intramolecular degrees of freedom, both suggested as possible origins of this process (see section 2.7). However, Fig. 7 documents that excellent fits of the broadband spectra of blood can be obtained assuming three distinct dispersion regimes only, namely electrode polarization, $\beta$, and $\gamma$ dispersion [24]. It is not necessary to assume the existence of a further $\delta$ relaxation, located between the $\beta$ and $\gamma$ processes to describe the smooth transition between these two processes which is simply caused by a bare superposition of the wings of the two neighboring processes. Thus, it was concluded that the $\delta$ dispersion observed in many protein solutions is absent in blood or at least strongly superimposed by the other contributions [24].

As discussed in section 2.4, the $\alpha$ process, representing the lowest-frequency quasi-intrinsic dispersion regime sometimes reported in tissues, is often assumed to arise from counterion diffusion along cell membranes [1,2,12,13,27] (but also other origins were discussed, see, e.g., [2,28,153]). Thus, it may also occur in blood, and, indeed, strong low-frequency dispersion is revealed in its spectra (Fig. 7). However, Wolf *et al*. [24] attributed this dispersion regime solely to electrode polarization (section 2.3). As this dispersion is also detected in blood plasma where RBCs are completely absent, it is quite evident that this low-frequency relaxation of whole blood is not related to the presence of any dynamics at the cell membranes.

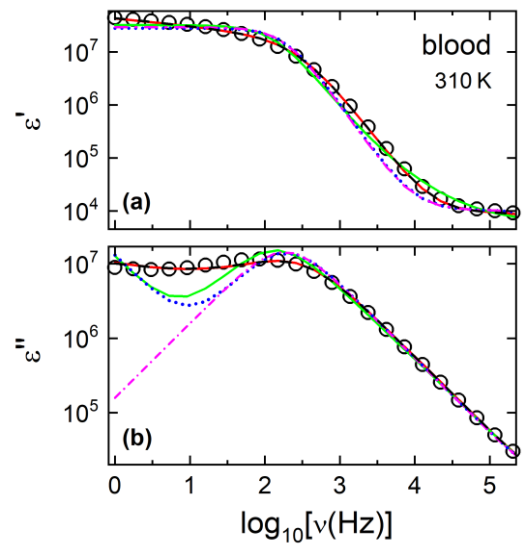

**Fig. 8** Frequency dependence of dielectric constant (a) and dielectric loss of whole blood at low frequencies and body temperature (reprinted from Ref. [24] with permission from Elsevier). The lines represent fits using a Debye function (dash dotted magenta line), a Debye (dotted blue line) and CC function (green line), both with additional dc-conductivity, a distributed RC equivalent circuit (red line), as well as a constant phase element (black dashed line).

To support the interpretation in terms of electrode polarization, in Ref. [24] a careful analysis of the low-frequency dispersion of whole blood was provided. As demonstrated in Fig. 8 by the dash-dotted magenta line, the low-frequency dispersion clearly cannot be fitted by a Debye relaxation. Even the assumption of additional dc conductivity (blue dotted line) cannot explain the shallow minimum appearing towards the lowest frequencies. Identical frequency dependence as the dotted line also results when assuming a simple RC equivalent circuit in series to the bulk sample, as is often done to describe canonical Maxwell-Wagner effects of inhomogeneous matter [18,158]. Thus, obviously such an approach also cannot account for the measured data. Substituting the Debye model by a CC function (green solid line) also does not lead to a reasonable fit of the spectrum. Good fits, however, were achieved, assuming a distributed RC circuit [18] (red dashed line) or a constant phase element (blue dashed line), both models pointing towards the existence of electrode polarization, which describes blocking electrodes in ion conductors (see section 2.3). Wolf *et al.* [24] concluded that in broadband dielectric spectra on blood there is no evidence for $\alpha$ dispersion due to counter-ion diffusion as sometimes reported for tissues. However, they also stated that a weak $\alpha$ relaxation cannot be fully excluded to be hidden by the other dominating contributions in the spectra.

## *3.4 THz Spectra of Amino Acids*

Amino acids are the fundamental building blocks of proteins and, hence, are essential for the evolution of life on Earth. They are linked together to polypeptide chains, which finally are folded into proteins. Amino acids are defined as molecules forming a chain of at least two carbon atoms, one belonging to an acidic carboxy, and the second to a basic amino group. In $\alpha$ amino acids, these two carbon atoms are directly linked together. In aqueous solutions, amino acids exist in their zwitterionic form with a net proton transfer from the carboxy to the amino group. $\alpha$ amino acids have a common structure, consisting of a central $\alpha$ carbon atom, connected to a charged amino group ($NH^{3+}$), a carboxyl group ($COO^-$), a hydrogen atom, and a variable organic side chain *R*. The side group determines the different chemical properties and the specific biological role of each amino acid. 22 $\alpha$ amino acids are genetically encoded (proteinogenic) with 20 of them belonging to the standard genetic code. Interestingly, despite their importance for the evolution of life, it still is a matter of controversy whether they were synthesized on primitive Earth or were transported via meteorites to our home planet [159,160].

Terahertz radiation, covering a frequency regime from 0.1 to 10 THz, corresponding to ~3–300 cm$^{-1}$, defines the long wavelength part of the infrared (IR) spectral range partly overlapping with the far-IR (FIR) regime. For a long time, this frequency range was only accessible by elaborate backward-wave oscillator 27techniques [161] in the laboratory and, hence, was termed "terahertz gap". Approximately during the past twenty years, based on the time-resolved detection of ultrashort radiation bursts of electric fields by photoconductive switches, this technique has undergone an outstanding development, nowadays known as terahertz time-domain spectroscopy (THz-TDS) [162], and plays a dominant role in optical spectroscopy in a variety of different fields in condensed matter physics as well as in material and biological sciences. In optical spectroscopy on molecular crystals, it is especially sensitive to intermolecular excitations (see section 2.8).

Vibrational spectroscopy on crystalline amino acids routinely was performed utilizing Raman and IR spectroscopy, and inelastic neutron scattering techniques. Infrared spectroscopy on amino acids started with the work of Klotz and Gruen [163] and Thompson *et al*. [164]. Raman spectroscopy started even earlier and was initiated by a series of publications by Edsall and co-workers [165,166]. Terahertz spectroscopy on biological samples started in the early 21st century and largely helped to explore the hardly accessible frequency range between the FIR and the microwave regime [167,168]. In addition, THz-TDS with its ability to concomitantly monitor real and imaginary parts of the optical constants and its high transmission through the material, combined with little damage, is of utmost importance for the investigation of biological matter.

Recent THz work on the amino acids L-serine (Ser) and L-cysteine (Cys) in the frequency regime from 40–120 cm$^{-1}$ was published by Emmert *et al*. [11], with a detailed analysis of the temperature dependence of eigenfrequencies, damping, and dipolar strength, and including a complete reference list of Raman, FIR, and THz results specifically with reference to Cys and Ser. In addition, in Ref. [11] by referring to existing model calculations and computer simulations, it was attempted to assign the observed eigenfrequencies to the microscopic nature of specific vibrational modes. Here we refer to this work and add unpublished THz results of L-asparagine (ASN; $C_4H_8N_2O_3$). Experimental details can be found in Ref. [11].

L-asparagine has a side chain R=$CH_2CONH_2$, and, hence, this molecule contains four carbon atoms, with the terminating carbon in the side chain bound to oxygen and a $NH_2$ group. L-asparagine crystallizes in the monocline space group P2$_1$ with two molecules per unit cell ($M$ = 2) [169]. The single molecule exhibits 3$N$ - 6 modes (see section 2.8). In crystalline form one expects a total of 3$MN$ - 3 external plus internal modes. That is, in the case of ASN with 17 atoms per molecule and two molecules per unit cell, 99 modes are expected. The lattice constants are $a$ = 5.0622 Å, $b$ = 6.7001 Å, and $c$ = 8.0543 Å, with an angle between ($a,c$) $\beta$ = 91.706. The volume of the unit cell at room temperature $V$ = 273.06 Å$^3$ [169]. Vibrational spectroscopy utilizing FIR and THz, Raman, as well as inelastic neutron scattering (INS) results on ASN were reported in Refs. [170,171,172,173,174,175,176,177]. The measurements were mostly performed at room temperature only, sometimes complemented by model calculations, and, in some cases, attempts were made to assign specific modes to characteristic excitation patterns. We think that an interpretation of the microscopic displacement patterns of observed eigenfrequencies can best be made following the temperature dependence of the eigenmodes and, hence, we recorded the detailed temperature evolution of the real and imaginary parts of the complex dielectric constant.

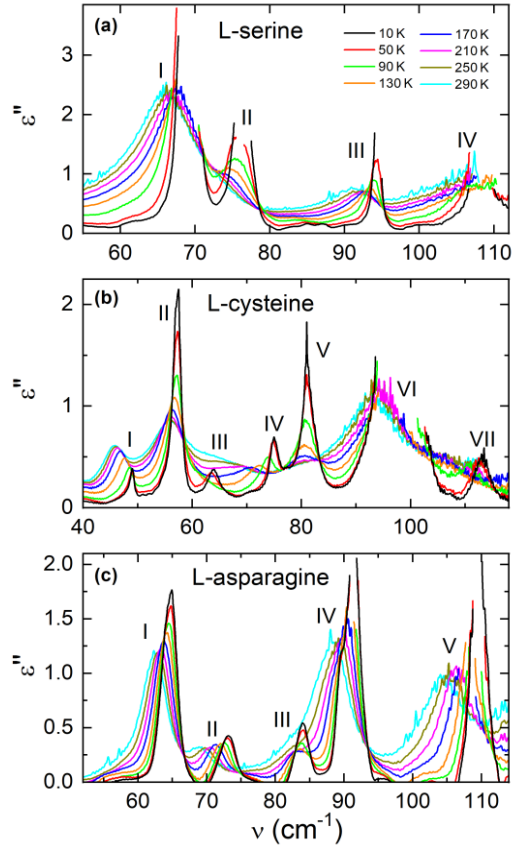

**Fig. 9** THz data of the wavenumber/frequency dependence of the dielectric loss of L-serine (**a**), L-cysteine (**b**), and ASN (**c**) as function of temperature between 10 and 290 K. The sequence of eigenmodes on increasing frequency is labelled by Roman numerals. For the strongest peaks, close to resonance, data points are missing due to too high absorption for the given sample thickness (see text). The results on L-serine and L-cysteine were taken from Ref. [11].

In the forthcoming figures we discuss the temperature dependence of the dielectric loss $\varepsilon''$, which seems to be most informative with respect to eigenfrequencies and widths of the involved excitations. Figure 9 shows the temperature dependence of the dielectric loss of ASN in a wavenumber regime from 45–115 cm$^{-1}$ [Fig. 9(c)], compared to the results obtained in L-serine (a) and L-cysteine (b), the latter being published in Ref. [11]. In good approximation, the eigenfrequencies of the observed vibrations correspond to the center of the Lorentzian peaks. In all spectra, mainly at low temperatures, experimental data points in the center of the most intensive peaks are missing, due to high absorption in the investigated samples. In these experiments, we had to make a compromise and tried to choose the sample thickness in a way that allows measurements of strong and weak intensities concomitantly at low as well as at high temperatures. With respect to the crystallographic structure, ASN crystallizes in a monoclinic space group, while Ser and Cys are orthorhombic and, hence, one might expect similarities in the spectra of Ser and Cys, different to those observed in ASN. However, experimentally the frequency dependencies of Cys and ASN reveal astonishing similarities, with the sequence of modes II, III, IV, V, and VI of Cys, recaptured in modes I, II, III, IV and V in ASN, with the latter three modes shifted to higher frequencies.

The spectra of all amino acids, investigated in the course of this work and documented in Fig. 9, reveal well-defined Lorentzian peaks, which undergo a red shift and concomitant broadening on increasing temperatures, canonical anharmonic effects as observed in many condensed-matter systems. Ser and Cys were discussed in some detail by Emmert *et al.* [11]. Specifically, it remained unclear if the very weak absorption features detected in Ser close to ~ 62, 86, and 100 cm$^{-1}$ at the lowest temperatures correspond to weak intrinsic excitations, result from impurities, or are due to background scatter. Focusing on ASN, in the frequency regime investigated, we observe five well-defined excitations at low temperatures, all of them showing considerable red shift and broadening on heating. However, most of the peaks are still well defined at room temperature, only with mode III being slightly overdamped, appearing as hump in the low-frequency wing of mode IV. Mode IV possibly represents a double peak, best visible in the 10 K spectra, which, due to experimental uncertainties, could not be finally clarified.

For a more systematic analysis, we performed Lorentzian fits to the observed spectra, with eigenfrequency $\nu$, damping $\gamma$ (half width at half maximum), and optical weight $\Delta$, which characterizes the dipolar strength (for more details on the Lorentzian model used and on the fitting procedure, please see Ref. [11]). We always concomitantly fitted real (not shown) and imaginary part of the dielectric constant. Results of these fits are indicated in Fig. 10 by solid lines: As representative examples, we provide the frequency dependences of the dielectric loss at three temperatures. As documented in Fig. 10, the fits work reasonably well, with deviations mainly in the low-loss regions of the spectra, close to the background contributions characterized by low absorption. The observed red shifts and increasing dampings are well described, and at room temperature mode III is only visible as shoulder in the low-frequency wing of mode IV. In the temperature and frequency regime investigated, all the modes seem to be well behaved and their red shifts and increasing line widths on increasing temperature signal canonical anharmonic behavior of these modes as routinely observed in crystalline materials.

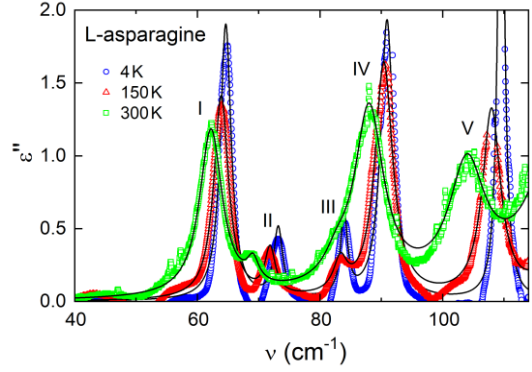

**Fig. 10** Temperature dependence of the dielectric loss $\varepsilon''$ of ASN as function of wavenumber $\nu$ at three selected temperatures 4, 150, and 300 K. The lines represent the concomitant fits of the real (not shown) and imaginary parts of the dielectric permittivity using Lorentzian line shapes. Deviations in the low-loss regions result from background scatter of the polycrystalline samples.

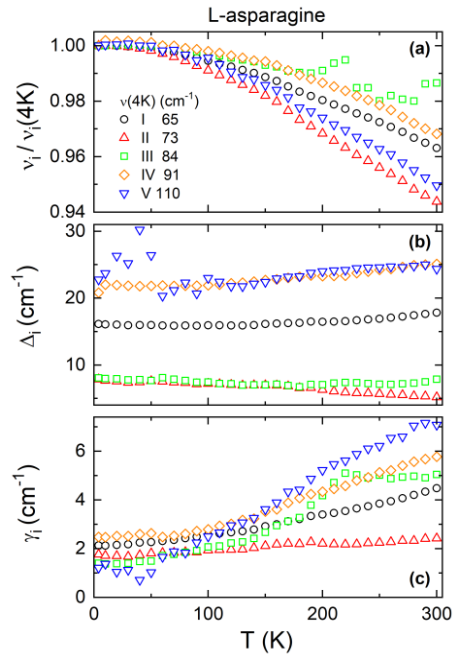

**Fig. 11** Temperature dependence of (**a**) the normalized eigenfrequencies $\nu_i/\nu_i(4\ \text{K})$, (**b**) the oscillator strengths $\Delta_i$, and (**c**) the damping constants $\gamma_i$ derived from the Lorentzian fits of the five modes observed in ASN.

In Fig. 11, we document the results of these fits, namely temperature dependencies of the normalized eigenfrequencies [$\nu(T)/\nu(4K)$], of the optical weight or dipolar strength $\Delta(T)$, and of the damping of the eigenmodes $\gamma(T)$. In condensed-matter systems, we expect a continuous increase of the eigenfrequencies with decreasing $T$ due to anharmonic effects, saturating towards low temperatures and exhibiting a linear decrease towards temperatures that are higher than the Debye temperature. Concomitantly, we expect a similar increase in the temperature-dependent damping of the eigenmodes. The dipolar strength or optical weight mainly depends on the static or dynamic dipole moments of the involved vibrations and, in first approximation, is expected to be temperature independent. Figure 11 reveals that this expected canonical anharmonic behavior describes the experimentally observed results reasonably well. The observed scatter in the data, specifically observed for modes III and IV, results from experimental uncertainties. Figure 11(a) shows normalized frequencies only, and the characteristic eigenfrequencies at 4 K as indicated in this frame are: 65, 73, 84, 91, and 110 cm$^{-1}$.

Finally, by comparing these results to existing model calculations as well as to published results on vibrational excitations in ASN, we would like to gain some insight into the microscopic nature of the observed excitations. First, the continuous temperature evolution of the relevant parameters of the eigenmodes signals that in ASN there is no phase transition at low temperatures (< 300 K), as was observed e.g., in L-cysteine [11]. In literature, there are several published data sets of excitations in ASN utilizing FIR, Raman, and INS techniques, in some cases complemented by model calculations [170,172,173,175,176,177]. Most of the published experiments focus on excitations > 100 cm$^{-1}$ and only a few experimental results of excitations with wavenumbers < 110 cm$^{-1}$ are available. Matei et al. [172] reported FIR spectra of ASN at room temperature and found weak absorption lines at 62, 87, and 103 cm$^{-1}$ coming close to our results of modes I, IV, and V. As documented in Fig. 9, these are the strongest peaks in our frequency regime and Matei et al. [172] obviously missed the weak lines denominated as mode II and III in our work. These authors ascribed their observed excitations to hydrogen-bond modes. In this frequency regime and at room temperature, Sylvestre et al. [175] reported the observation of two modes by FIR techniques at wavenumbers of 68 and 84 cm$^{-1}$, probably corresponding to modes I and IV, however, with relatively large uncertainties. These authors describe the 68 cm$^{-1}$ excitation to a backbone bending of the carbon atoms coupled to a torsional motion of some end groups. By performing detailed INS and FIR experiments, Pawlukoj et al. [176] reported eigenfrequencies detected by IR techniques at room temperature at 62 and 88 cm$^{-1}$ and by INS at 30 K at 75 and 87 cm$^{-1}$. The high-frequency excitation probably corresponds to our mode IV, which they ascribe to a torsional motion of the $CO_2^-$ group, and the low-frequency mode at 62 cm$^{-1}$ reported from their FIR experiments, which comes close to our mode I, they assign to lattice vibrations, without further specification. The mode they observe by INS at 30 K, cannot be assigned in our THz experiments. Navarette et al. [178] studied the excitation spectra of ASN in solution. They calculated all the eigenfrequencies of the molecule, and found the three lowest excitations at 48, 84, and 91 cm$^{-1}$, which is only partly compatible with our experimental results. They ascribe these eigenfrequencies mainly to torsional and bending excitations of the carbon backbone and the carboxy group. This discussion and comparison to other work makes clear how difficult it is to assign modes, specifically at low wave numbers, to identify the nature of a specific molecular excitation.

The similarity of the spectra of ASN and L-cysteine mentioned before makes it reasonable to identify most of these modes resulting from coupled intermolecular translational and librational excitations of the rigid amino acid molecules. This certainly coincides with a detailed INS study of single-crystalline L-alanine by Micu *et al*. [179]. These authors document that the vibrational spectra below 100 cm$^{-1}$ are dominated by intermolecular translational optical and librational modes of the center of mass of the molecules (see section 2.8). However, these excitations certainly are strongly coupled with rotational excitations of the backbone, of the side chain or of molecular side groups.

# 4 Concluding Remarks

As discussed in the introduction, performing dielectric spectroscopy on biological matter can reveal information that not only is of interest from an academic point of view but also highly valuable for applications in the fields of dosimetry, diagnostics, and therapeutic medicine. Arriving at a proper microscopic understanding of such spectra, thus certainly is an important task. However, the overview of the many dynamic processes that can contribute to the dielectric response of biological matter, provided in section 2 of the present work, and the experimental examples discussed in section 3 demonstrate that this task can be challenging. Even for such simple molecules as water (section 3.1), several different dynamics must be considered to understand the experimental results, and, despite many-decades of research efforts, no consensus is reached about their correct interpretation. For larger biological molecules, the situation usually becomes worse. For example, the amino-acid investigations treated in the present work (section 3.4) show that, even in their solid form and in the restricted THz frequency range dominated by resonances only (section 2.8), the behavior can be very complex and the interpretation quite demanding. Dissolving biological macromolecules like proteins in water, which comes closer to in vivo conditions, generates various additional dispersion effects, especially in the classical dielectric-spectroscopy frequency range (section 3.2): Aside from the tumbling dynamics of the dipolar bio- and water molecules (sections 2.5 and 2.6), further processes can be expected, arising, e.g., from bound water and from intramolecular degrees of freedom of macromolecules, both considered to be important for the functionality of proteins (section 2.7). As the aqueous solvent usually contains considerable amounts of ions, significant spectral contributions from electrode polarization can arise at low frequencies (section 2.3), which should not be mistaken for intrinsic dynamics. In the dielectric constant, they strongly superimpose the other dynamics of interest, analogous to the influence of the dc conductivity on the low-frequency dielectric loss (section 2.2), which also can obscure intrinsic spectral features. Thus, only careful analysis can unveil the true dielectric behavior at low frequencies. For actually functional biological matter in organisms, like tissues or blood (section 3.3), the presence of cells introduces mesoscopic inhomogeneities in the samples that can give rise to various additional relaxational features in the dielectric spectra. They partly have to be considered as essentially non-intrinsic, arising from heterogeneity-triggered Maxwell-Wagner mechanisms (section 2.5). As then the frequencies and other parameters of the associated loss peaks do not necessarily reflect the time scales of any microscopic motions within the material, only a proper evaluation in terms of model calculations can reveal the intrinsic properties.

Listing up some of the most important possible mechanisms that were considered to manifest in dielectric spectra of biological matter, we have: electrode polarization, ionic dc conductivity in the solvent and within the cellular cytoplasm, counter-ion diffusion along cell membranes causing the elusive $\alpha$ relaxation, conformational sampling motions (sub-$\beta$), reorientations of dipolar macromolecules like proteins ($\beta$ relaxation), Maxwell-Wagner relaxation of cell membranes ($\beta$), bound-water dynamics (maybe bimodal due to loosely and tightly bound layers; $\delta$ relaxation), collective protein-water motions ($\delta$), intramolecular motions of macromolecules ($\delta$), unbound-water reorientations ($\gamma$), and phonon-like or vibrational resonances of intra- and intermolecular origin. Here we omitted the possible additional secondary dynamics like the excess wing [14,15,111] or the Johari-Goldstein $\beta$ relaxation [180], and the boson peak [14,74,75,76,77], all mainly known from the field of disordered materials such as glass forming liquids. However, generally there is no reason to dismiss their existence in biological matter, which usually also is disordered.

The above list is quite overwhelming, especially when considering that, in principle, all of these processes can exist in tissues and blood, many of them in solutions of macromolecules, and at least several of them in simpler molecules like water or solid amino acids. Nevertheless, even for dissolved proteins, tissues, or blood, the experimentally determined dielectric spectra are astonishingly simple. Obviously, the spectral features caused by many of the listed processes are absent or totally obscured by those from other, stronger processes and, thus, undetectable in the spectra. The remaining, detectable ones often are superimposed to each other. Therefore, to perform meaningful dielectric investigations of biological materials, three requirements should be fulfilled: (i) A sufficiently broad frequency range should be covered by the experiments. To obtain correct parameters for the dynamics of interest, then one is able to correctly account for overlapping contributions from the neighboring processes in the fits of the data. (ii) A proper modeling of the experimental results employing: suitable equivalent circuits, the broadened relaxation functions known from glass physics (instead of multiple Debye functions, which often is misleading), and the models developed for the analysis of apparent relaxations arising from the presence of suspended cells. (iii) The collection of dielectric spectra in dependence of various parameters, such as temperature, sample geometry, cell concentration, etc. This can help to discriminate between alternative origins of a detected relaxation feature.

We think the present work demonstrates that complying with these requirements can indeed help in drawing significant conclusions from broadband dielectric and THz measurements on bio-related matter, and we hope it will further advance the art of performing proper dielectric investigations of this class of materials.